\documentclass[aps,twocolumn,pra,superscriptaddress]{revtex4}

\usepackage{amsmath}
\usepackage{amsthm,comment}
\usepackage[colorlinks=true,citecolor=blue,linkcolor=red,urlcolor=red]{hyperref}
\usepackage{graphicx}
\usepackage{braket}
\DeclareGraphicsExtensions{.pdf}

\newcommand{\beq}{\begin{equation}}
\newcommand{\eeq}{\end{equation}}
\newcommand{\bea}{\begin{eqnarray}}
\newcommand{\eea}{\end{eqnarray}}

\newcommand{\eu}{\mathrm{e}}

\newcommand{\dg}{^\dagger}

\newcommand{\const}{\mathrm{const}}

\newcommand{\q}{{\bf q}}

\newcommand{\Ho}{{\hat H}}

\newcommand{\be}{\beta}

\newcommand{\si}{\sigma}

\newcommand{\psio}{\hat{\psi}}
\newcommand{\xv}{{\bf x}}
\newcommand{\xvo}{{\hat\xv}}
\newcommand{\pv}{{\bf p}}
\newcommand{\pvo}{{\hat\pv}}
\newcommand{\rv}{{\bf r}}
\newcommand{\sv}{{\bf s}}
\newcommand{\Jvo}{{\hat{\bf J}}}

\newcommand{\kv}{{\bf k}}
\newcommand{\D}{\operatorname{d\!}}

\newcommand{\Dc}{\mathcal{D}}
\newcommand{\ro}{{\hat\rho}}
\newcommand{\rod}{{\hat\varrho}}
\newcommand{\rodk}{\rod_\kv}
\newcommand{\rodmk}{\rod_{-\kv}}
\newcommand{\Jvok}{\Jvo_\kv}
\newcommand{\Jvomk}{\Jvo_{-\kv}}

\newcommand{\Lok}{\hat L_\kv}

\newcommand{\Lvo}{\hat{\bf L}}
\newcommand{\Lo}{\hat L}
\newcommand{\Tb}{T_\be}
\newcommand{\Esi}{E_\si}
\newcommand{\kB}{k_\text{\tiny B}}

\begin{document}

\title{Linear friction many-body equation for dissipative spontaneous wavefunction collapse}

\author{Giovanni Di Bartolomeo}
\affiliation{Department of Physics, University of Trieste, Strada Costiera 11, 34151 Trieste, Italy}
\affiliation{Istituto Nazionale di Fisica Nucleare, Trieste Section, Via Valerio 2, 34127 Trieste, Italy}
\author{Matteo Carlesso}
\email{m.carlesso@qub.ac.uk}
\affiliation{Centre for Quantum Materials and Technologies, School of Mathematics and Physics, Queen’s University Belfast, BT7 1NN Belfast, UK}
\author{Kristian Piscicchia} 
\affiliation{Centro Ricerche Enrico Fermi -- Museo Storico della Fisica e Centro Studi e Ricerche ``Enrico Fermi'', Piazza del Viminale 1, 00184 Rome, Italy}
\affiliation{INFN, Laboratori Nazionali di Frascati, Via Enrico Fermi 54, 00044 Frascati, Italy}
\author{Catalina Curceanu}
\affiliation{INFN, Laboratori Nazionali di Frascati, Via Enrico Fermi 54, 00044 Frascati, Italy}
\author{Maaneli Derakhshani} 
\affiliation{Department of Mathematics, Rutgers University, 110 Frelinghuysen Road
Piscataway, NJ 08854-8019, USA}
\author{Lajos Di\'osi}
\affiliation{Wigner Research Center for Physics, H-1525 Budapest 114 , P.O.Box 49, Hungary}
\affiliation{E\"otv\"os Lor\'and University, H-1117 Budapest, P\'azm\'any P\'eter stny. 1/A}

\date{\today}
\begin{abstract}
We construct and study the simplest universal dissipative Lindblad master equation
for many-body  systems with the purpose of  a new dissipative extension of existing non-relativistic theories of fundamental spontaneous decoherence  and spontaneous wave function collapse in Nature. It is universal as it is written in terms of second-quantized mass density $\rod$ and current
$\Jvo$, thus making it
independent of the material structure and
its parameters.
Assuming a linear friction in the current, we find that the dissipative structure is strictly
constrained. Following the general structure of our dissipative Lindblad equation, we derive and analyze the dissipative extensions of the two most known spontaneous wave function collapse models, the  Di\'osi-Penrose and the Continuous Spontaneous Localization models.
 \end{abstract}

\maketitle

\section{Introduction}\label{Intro}
Testable predictions of quantum theory assume the presence of measuring devices
providing data on the quantum system in question. This process, called quantum measurement, yields the collapse of superpositions
in accordance with random measurement outcomes. 
The concept of spontaneous collapse can be interpreted as the hypothesis that measurements are occurring spontaneously at each point of space and  time according to some universal protocol, but without the presence of actual measuring devices \cite{Dio18}. For a comprehensive review of the theory of spontaneous collapse models and their experimental testing,  
see reviews \cite{BasGhi03,Basetal13, natphysKlaus, NatPhysMatteo}. 
Through a suitable choice of its structure and parameters, the protocol ensures that
such spontaneous collapses of the wavefunctions $\phi$ become significant for macroscopic (e.g., massive) 
systems but remains negligible for microscopic ones. Currently, two
non-relativistic models, the Di\'osi-Penrose (DP) and the Continuous Spontaneous Localization (CSL), have been crystallized \cite{DP1, DP2,CSL1, CSL2}, 
and they correspond to the spontaneous measurement of the mass spatial density operator
$\rod(\rv,t)$ at all $\rv$ and $t$. Correspondingly, the persistence of (macroscopic) superpositions is
lost and the unitary dynamics is modified. Once the statistical average is considered, the corresponding dynamical equation for the \textit{statistical operator} $\hat\rho$ describing such models is a Lindblad master equation borrowed from open quantum systems theory.
The noise generated by such a dynamics leads to a low-rate
spontaneous heating. Nonetheless, the accumulation of
such a heat is problematic, even for a phenomenological model.
The basic form of the DP and the CSL models'  master equations lead to decoherence without
dissipation. To dissipate the spontaneously generated heat, a dissipative mechanism
needs to be included to the basic master equations.  Attempts in such a direction were made \cite{smirne2015dissipative,BahSmiBas14, Gaida} and some experiments were used to test the theory   \cite{dissinterf, unitary,Pontindiss,vinantediss}.

Here, we consider the simplest
many-body dissipative Lindblad master equation where the friction term is
linear in the second-quantized current $\Jvo(\rv,t)$. We show that  our choice of the Lindblad collapse operator, which is independent from the details of the considered system, leads to the dissipation of the current and of the mean energy. We construct and analyze the corresponding dissipative extensions of the DP and CSL models.

\section{Spontaneous decoherence models}\label{DPCSL}
We start from a modified von-Neumann--Schr\"odinger (master) equation
\beq\label{ME}
\dot\ro=-\frac{i}{\hbar}[\Ho,\ro]+\Dc\ro,
\eeq
where $\Ho$ is the many-body Hamiltonian describing the Schr\"odinger dynamics, and $\Dc$ is a new term introducing the action of spontaneous decoherence. For the latter, we consider a simple 
Lindblad form corresponding to the spontaneous measurement of the second-quantised mass density $\rod(\rv)=m\psio\dg(\rv)\psio(\rv)$ with $\psio(\rv)$ being the (fermionic) annihilation field-operator. Explicitly, it  reads
\bea\label{D}
\Dc\ro&=&-\frac{1}{2\hbar^2}\int\int \D\rv \D\sv D(\rv-\sv)[\rod(\rv),[\rod(\sv),\ro]],\nonumber\\
             &=&\frac{1}{\hbar^2}\int \frac{\D\kv}{(2\pi)^3} D_\kv
            \left(\rodk\ro\rodk\dg-\tfrac12\{\rodk\dg\rodk,\ro\}\right),
\eea
where we introduced the Fourier transform of the mass density $\rodk$ and the kernel $D$. Depending on the explicit space (or momentum) dependence of the kernel, one might want to introduce a  short-length regularization, typically in the form of a Gaussian smearing of the field $\rod(\rv)$. 
In the Fourier representation, a Gaussian smearing of scale $\si$ takes a simple form for both the models (DP and CSL) we will consider: 
\beq\label{Dk}
D_\kv=\exp\left(-\si^2 k^2\right)\times\left\{\begin{array}{lll}
4\pi\hbar G/k^2&\mbox{ (DP)}\\ 
\hbar^2\gamma&\mbox{ (CSL)}
\end{array}\right.
\eeq
In the DP model, the decoherence rate is set by the Newton constant $G$ and
the kernel contains a $1/k^2$ factor in addition to the smearing prefactor. Using an alternative notation, $R_0=\si$ is the spatial cutoff in the DP model and it is the only free parameter of the model. Current experimental bounds set the typical smearing at subatomic length-scales $\si\geq 5\times 10^{-11}\,$m \cite{Donetal21}, although larger values can be also considered \cite{Saletal08, spaceinterf}.
On the other hand, the CSL model can be  described in terms of two free parameters being $\lambda=\gamma m_0^2/(\sqrt{4\pi}r_\text{\tiny C})^3$ and $r_\text{\tiny C}=\si$, which are respectively the collapse rate and localization length of the model ($m_0$ is a reference mass chosen as that of a nucleon). Conversely to the DP model, the typically considered values of the spatial smearing are around $\sigma\simeq10^{-7}\,$m, well in the mesoscopic regime.
A new mapping between the two models can be introduced, and it is based on the simple 
relationship  
$D_\text{\tiny CSL}=-\const\times\partial D_\text{\tiny DP}/\partial(\si^2)$
between the two kernels.
Consequently, the  decoherence term of the
CSL model can be obtained from the  DP one through
\beq\label{DrhoDPvCSL}
\Dc_\text{\tiny CSL}=-\frac{\hbar\gamma}{4\pi G}\frac{\partial\Dc_\text{\tiny DP}}{\partial(\si^2)}.
\eeq
The inverse integral relation can be also simply derived.
We anticipate that
these relations will survive in the forthcoming dissipative generalization
of the dissipator $\Dc$.

\section{Spontaneous heating}\label{heating}
The smaller the cutoff $\sigma$, the larger the strength of the  collapse
effect and the spontaneous heating as well \cite{Adler2021}. The latter implies a continuous
increase of the kinetic energy for each particle. We elucidate this mechanism on
a single point-like, free particle of mass $m$ and canonical variables $\xvo$ and $\pvo$. The corresponding  heating power $P$, i.e.~the time-derivative of the kinetic energy 
$\Ho=(\pvo^2/2m)$, is obtained from the master equation \eqref{ME}:
\beq
P=\frac{\D\braket\Ho}{\D t}=\braket{\Dc\dg\Ho}=\frac{1}{2m}\braket{\Dc\dg\pvo^2},
\eeq
where $\Dc\dg$ is given by
\beq\label{eq.adjoint}
\Dc\dg\hat O
=\frac{1}{\hbar^2}\int \frac{\D\kv}{(2\pi)^3} D_\kv
\left(\rodk\dg\hat O\rodk-\tfrac12\{\rodk\dg\rodk,\hat O\}\right).
\eeq
For the case under study, the mass density and its Fourier transform read
\beq\label{m}
\rod(\rv)=m\delta(\rv-\xvo),~~~~~\rod_\kv=m e^{i\kv\xvo}.
\eeq
We insert $\rod_\kv$ in Eq.~\eqref{eq.adjoint} and obtain the expression of the heating power
\beq
P=\frac{m}{2}\int \frac{\D\kv}{(2\pi)^3} D_\kv k^2.
\eeq 
where we used the identity 
\beq\label{Weyl}
e^{-i\kv\hat{\xv}}f(\hat{\pv})e^{i\kv\xvo}=f(\pvo+\hbar\kv),
\eeq
and the spherical symmetry of $D_\kv$. 
The integral is characteristic for the regularized behaviour of the kernel 
and we calculate it from the  Eq.~\eqref{Dk} for both models:
\bea\label{Dprimeprime}
P=-\frac{m}{2}D^{\prime\prime}(\rv)|_{\rv=0}=\frac{1}{\sqrt{(2\pi)^3}}\left\{\begin{array}{lll}
4\pi\hbar G/\si^3&\mbox{ (DP)}\\ 
3\hbar^2\gamma/\si^5&\mbox{ (CSL)}
\end{array}\right.
\eea
Hence, we get the following heating powers:
\begin{subequations}\label{eq.P.P}
\bea\label{PDP}
    P^\text{\tiny DP}&=&\frac{\hbar G m}{4\sqrt{\pi}\si^3},\\
    \label{PCSL}
    P^\text{\tiny CSL}&=&\frac{3m \hbar^2 \gamma}{32\pi^{3/2}\si^5},
\eea
\end{subequations}
which are
related to each other in conformity with the  mapping of Eq.~\eqref{DrhoDPvCSL}. 

To get an insight into the underlying effective mechanism, we consider the
dynamics of the momentum $\pvo$ in details, i.e.~the dynamics of arbitrary 
functions $f(\pvo)$ of momentum. 
It could be shown that the Heisenberg equation of motion of $f(\pvo)$ is closed, 
this important feature has its parallel in the equivalent von-Neumann--Schr\"odinger dynamics \eqref{ME} of the state $\ro$.
Insofar, since we are not interested in the dynamics of the coordinate $\xvo$ but
of $\pvo$, we can start with the specific form $\ro=\rho(\pvo)$ of the
state, then consider its evolution under the Fourier representation
of the dissipator in Eq.~\eqref{D}. The specific form $\rho(\pvo)$, diagonal
in momentum basis,  is preserved: 
\beq\label{}
\frac{\D\rho(\pvo)}{\D t}
=\frac{m^2}{\hbar^2}\int \frac{\D\kv}{(2\pi)^3} D_\kv
\Bigl(\rho(\pvo-\hbar\kv)-\rho(\pvo)\Bigr).
\eeq
The result is a semi-classical single particle kinetic equation. 
The effect of dissipator $\Dc$
is equivalent to the random jumps $\pv\rightarrow\pv+\hbar\kv$ in 
momentum at the isotropic probability rate
\beq\label{collrate}
\frac{m^2}{\hbar^2}\frac{\D \kv}{(2\pi)^3}D_\kv.
\eeq
Since the kernel $D_\kv$ contains the regularizing factor $\exp(-\si^2 k^2)$,
the elementary momentum and energy transfers are in a bounded range.
We introduce the characteristic bound of the elementary energy transfer, which reads
\beq\label{Esi}
\Esi=\frac{\hbar^2}{4m\sigma^2}. 
\eeq
The quantity $\Esi$ is important when, in the rest of our work, 
we balance the spontaneous heating by a new dissipative mechanism 
to reach a balance equation:
\beq\label{balance}
\frac{\D\braket\Ho}{\D t}=P-\Gamma\braket\Ho,
\eeq
with a dissipation rate $\Gamma>0$. In such a way, a finite asymptotic (equilibrium) energy
$\braket\Ho_\infty=P/\Gamma$ is reached for ${\D\braket\Ho}/{\D t}=0$, and the corresponding effective temperature $T$ is
defined by the equipartition theorem $\braket\Ho_\infty=\frac32\kB T$.  {We underline that the single free particle case is sufficient to show if the spontaneous heating effects lead to a divergence of the energy, or if they can be counterbalanced with a damping effect thus leading to a finite asymptotic energy. This is the same approach that was considered in \cite{smirne2015dissipative} for the CSL model and in \cite{BahSmiBas14} for the DP model.}

\section{Dissipative extension - an exercise}\label{exercise}
Before introducing our dissipative extension of the DP and the CSL models, it is instructive to understand the elementary Lindblad form of friction. 
We start by considering the master equation for a single free particle of Hamiltonian $\Ho=\pvo^2/2m$ with a decoherence term of the form 
\beq
\Dc\ro=-\frac{D}{\hbar^2}[\xvo,[\xvo,\ro]]. 
\eeq
This can be considered as a minimal 
model for the spontaneous measurement of $\xvo(t)$. It yields a spatial decoherence
at rate $D/\hbar^2$. 
Equivalently, it implies a momentum diffusion with diffusion constant $D/2$, yielding a 
constant heating at power $P=3D/m$.
A way to include dissipation is to replace the Hermitian Lindblad generator $\xvo$ with the non-Hermitian operator 
\cite{Dio95,BrePet02}:
\beq\label{Lxp}
\Lvo=\xvo+i\frac{\hbar\be}{4m}\pvo,
\eeq 
where $\be$ 
will turn out to be the inverse equilibrium temperature.
Correspondingly, the Lindbladian term of the master equation takes the form
\beq
\Dc\ro=\frac{2D}{\hbar^2}\left(\Lvo\ro\Lvo\dg-\tfrac12\{\Lvo\dg\Lvo,\ro\}\right).
\eeq
By expanding $\Lvo$, we find
\bea\label{Dxp}
&&\Dc\ro
=\frac{i}{\hbar}\left[D\frac{\beta}{4m}\{\xvo,\pvo\},\ro\right]-\\\nonumber
&&-\frac{D}{\hbar^2}\biggl(
[\xvo,[\xvo,\ro]]
+i\frac{\hbar\beta}{2m}[\xvo,\{\pvo,\ro\}]
+\frac{\hbar^2\beta^2}{16m^2}[\pvo,[\pvo,\ro]]
\biggr),
\eea
where we can cancel the first Hamiltonian term with a counterterm in $\Ho$.
The first term of the second line generates the heating power $P$ as before.
The second term yields to the standard mechanical friction: via the Heisenberg equation $\mathcal{D}^\dag\pvo=-\eta\pvo$ 
with friction coefficient $\eta=2\be D/m$. 
The second term imposes  a relaxation mechanism also for $\pvo^2$. For the mean
kinetic energy we get
\beq\label{dH}
\frac{\D\braket\Ho}{\D t}=\frac{3D}{m}-\frac{2\be D}{m}\braket\Ho.
\eeq
This is the balance expressed in Eq.~\eqref{balance} with power $P=3D/m$ and
positive dissipation rate $\Gamma=2\be D/m$. We get a finite equilibrium
energy $\braket\Ho_\infty=\tfrac32\be^{-1}$. We conclude that the effective
temperature is $T=1/(\kB\be)$.  
Fortunately, it is known that the equilibrium state of the 
master equation with the dissipator in Eq.~\eqref{Dxp} is the exact Gibbs state \cite{Dio95,BrePet02}:
\beq\label{gibbs}
\ro_\be=\mathcal{N}\eu^{-\be\pvo^2/2m},
\eeq
hence $1/(\kB\be)$ is not only an effective temperature but the true one.

\section{The many-body master equation of linear friction}\label{DPCSLdiss}
Now we introduce our model.
In analogy with the single particle linear friction in the previous section,  we replace
the Hermitian Lindblad generators $\rod(\rv)$ in the dissipator of Eq.~\eqref{D}
with a non-Hermitian operator   of the form
\beq\label{Lr}
\Lo(\rv)=\rod(\rv)-i\frac{\hbar\beta}{4}\nabla_\rv\Jvo(\rv),
\eeq
whose Fourier representation is
\beq\label{Lk}
\Lok=\rodk+\frac{\hbar\beta}{4}\kv\Jvok,
\eeq
and where we introduced the current
\beq\label{Jr}
\Jvo(\rv)=-i\frac{\hbar}{2}\left(\hat\psi\dg(\rv)\nabla_\rv\hat\psi(\rv)-\nabla_\rv\hat\psi\dg(\rv)\hat\psi(\rv)\right).
\eeq
We note that the anti-Hermitian part of $\Lo(\rv)$ needs to be a scalar, like the Hermitian
part $\rod(\rv)$. Indeed, one cannot just take the current $\Jvo(\rv)$, but needs to take its divergence. 
With this choice for the Lindblad generator, the master equation becomes
\bea
\Dc\ro\!&=&\!\frac{1}{\hbar^2}\!\!\int\!\!\!\int\!\! \D\rv \D\sv\, D(\rv-\sv)\!
                \left(\Lo(\rv)\ro\Lo\dg(\sv)\!-\!\tfrac12\{\Lo\dg(\sv)\Lo(\rv),\ro\}\right),\nonumber\\
                \label{compact.master}
\!&=&\!\frac{1}{\hbar^2}\!\!\!\int\!\! \frac{\D\kv}{(2\pi)^3}D_\kv
                \left(\Lok\ro\Lok\dg\!-\!\tfrac12\{\Lok\dg\Lok,\ro\}\right).
\eea
{
A possible unravelling of Eq.~\eqref{compact.master}, which provides the stochastic and non-linear dynamical equation for the wavefunction and describes its collapse, can be derived by following the prescription highlighted in Eq.~(5) of \cite{unitary}. Alternatively, one can construct such an unravelling starting from the structure in Eq.~(4) of \cite{smirne2015dissipative}}.
Merging the expression in Eq.~\eqref{compact.master}
 with the definition of $\hat L(\rv)$, we get a Hamiltonian term, which is suitably reabsorbed, and the following equivalent structures are obtained: 
\begin{widetext}
\bea\label{Ddiss}
\Dc\rod&=&-\frac{1}{2\hbar^2}\int\int \D\rv \D\sv \,D(\rv-\sv)
              \left([\rod(\rv),[\rod(\sv),\ro]]-\frac{i\hbar\be}{2}[\rod(\rv),\{\nabla_\sv\Jvo(\sv),\ro\}]+\frac{\hbar^2\be^2}{16} [\nabla_\rv\Jvo(\rv),[\nabla_\sv\Jvo(\sv),\ro]]\right),\nonumber\\
               &=&-\frac{1}{2\hbar^2}\int \frac{\D\kv}{(2\pi)^3} D_\kv 
             \left([\rodmk,[\rodk,\ro]]-\frac{\hbar\be}{2}[\rodmk,\{\kv\Jvok,\ro\}]+\frac{\hbar^2\be^2}{16} [\kv\Jvomk,[\kv\Jvok,\ro]]\right),
\eea
\end{widetext}
where the three  terms of  $\Dc$ are respectively responsible for the decoherence in mass density $\rod$, 
the damping of the current $\Jvo$ (dissipation), and the  decoherence in the (divergence) of the current $\Jvo$,
respectively.
The corresponding Heisenberg equation of motion for an arbitrary observable $\hat O$ can be obtained from the adjoint of the master equation $\dot{\hat O}=\frac{i}{\hbar}[\Ho,\hat O]+\Dc^\dagger\hat O$, where
\begin{widetext}
\begin{equation}    \label{Heis}
    \begin{aligned}
\Dc^\dag \hat O&=-\frac{1}{2\hbar^2}\int\int \D\rv \D\sv D(\rv-\sv)
\biggl([\rod(\rv),[\rod(\sv),\hat O]]
+\frac{i\hbar\be}{2}\{\nabla_\sv\Jvo(\sv),[\rod(\rv),\hat O]\}+\frac{\hbar^2\be^2}{16} [\nabla_\rv\Jvo(\rv),[\nabla_\sv\Jvo(\sv),\hat O]]\biggr),\\
&=-\frac{1}{2\hbar^2}\int \frac{\D\kv}{(2\pi)^{3}} D_{\kv}
\biggl([\rod_{-\kv},[\rod_{\kv},\hat O]]
+\frac{\hbar\be}{2}\{\kv\Jvo_{\kv},[\rod_{-\kv},\hat O]\}+\frac{\hbar^2\be^2}{16} [\kv\Jvo_{-\kv},[\kv\Jvo_{\kv},\hat O]]\biggr),
    \end{aligned}
\end{equation}
\end{widetext}

It is  central to this work to confirm that the second term in Eq.~\eqref{Heis} is indeed a damping of the current. 
The corresponding contribution to the evolution of $\Jvo(\rv)$ is given by
\beq\label{eq.J2}
{\dot\Jvo(\rv)}|_2=- \frac{i\be}{4\hbar}\int\!\!\int\!\! \D\sv \D\sv'\, D(\sv-\sv')\{\nabla_\sv\Jvo(\sv),[\rod(\sv'),\Jvo(\rv)]\}.
\eeq
We can calculate the commutator of the second quantized (fermionic) density
and current:
\beq
[\rod(\sv'),\Jvo(\rv)]=-i\hbar\nabla_{\sv'}\left[\delta(\sv'-\rv)\rod(\rv)\right].
\eeq
Inserting it in Eq.~\eqref{eq.J2} and integrating the latter by parts, we get
\beq
{\dot\Jvo(\rv)}|_2
=-\frac{\be}{4}\int \D\sv \nabla_\rv\circ\nabla_\sv D(\sv-\rv)\{\Jvo(\sv),\rod(\rv)\},
\eeq 
where $\circ$ indicates the tensor product.
We expand the anti-commutator of the fermionic density and current: 
\bea\label{eq.anti}
\tfrac12\{\rod(\rv),\Jvo(\sv)\}&=&m\delta(\rv-\sv)\Jvo(\rv)\\
&-&\frac{i\hbar}{2}m\left(\psio\dg(\rv)\nabla_\sv\psio\dg(\sv)\psio(\rv)\psio(\sv)
-\mathrm{H.C.}\right).
\nonumber
\eea
Using this, 
we find that
\beq\label{Jdot}
{\dot\Jvo(\rv)}|_2
=-\eta\Jvo(\rv)+
{i\left(\hat{\mathbf Y}(\rv)-\mathrm{H.C.}\right) }
\eeq
{where 
\beq
\hat {\bf Y}(\rv)=
\frac{
\beta\hbar m}{4}\!\!\int \D\sv \nabla_\rv\circ\nabla_\sv D(\sv-\rv)\psio\dg(\rv)\nabla_\sv\psio\dg(\sv)\psio(\rv)\psio(\sv)
.
\eeq
The latter contributes only when multiple fermions are present, while it vanishes when applied to a single fermion state. As a first order approximation, we neglect $\hat {\bf Y}(\rv)$ contributions.
}
Then, the current effectively decays with a friction rate
\beq
\eta=-\frac{\beta m}{2}D^{\prime\prime}(\rv)|_{\rv=0},
\eeq
which depends crucially on the parameter $\si$ 
that regularizes $D(\rv)$ at $\rv=0$, see  expressions in Eq.~\eqref{Dprimeprime}. In Appendix \ref{app.1}, we show that, in the case of a single particle, Eq.~\eqref{Jdot} holds with no approximations.

Since the methods to infer exact analytic features of the dissipative master equation are limited, we turn to the special single particle case. In such a case the mechanism of
dissipation is transparent, and exact analytic calculations are possible. {Moreover, 
in the case of both the standard CSL and DP models, i.e.~with no dissipation included, the heating rate is independent from the presence of interactions or external potentials.
This has been well addressed in \cite{tilloy}. The inclusion of dissipative effects however raptures this simple feature in dense interactive fermionic matter. To include interaction and fermionic exchange one should employ perturbative methods such those used in \cite{adler2019testing}, which can be applied independently of the Hamiltonian structure. However, this goes beyond the scope of the manuscript. Nonetheless, as long as our single fermion approximation is valid the thermodynamics remains trivial as in the non-dissipative case, while the  rates and the equilibrium temperature are calculable exactly as for single fermions.} 

\section{Single particle dissipative mechanism}\label{single}
In case of the single particle, it is most convenient to work in the Fourier representation of the mass density and the current:
\beq
\rodk=m e^{i\kv\xvo},~~~~\Jvok=\tfrac12\{\pvo,e^{i\kv\xvo}\}.
\eeq
Then, the Lindblad generator, leading  to the many-body dissipator $\Dc$ 
shown in Eq.~\eqref{Ddiss},
reduces to simple alternative forms:
\bea\label{Lok1}
\Lok
&=&\left(m-\frac{\hbar^2 k^2\be}{8}+\frac{\hbar\kv\be}{4}\pvo\right)\eu^{i\kv\xvo},\nonumber\\
&=&\eu^{i\kv\xvo}\left(m+\frac{\hbar^2 k^2\be}{8}+\frac{\hbar\kv\be}{4}\pvo\right).
\eea
Let us see how the kinetic equation of the momenta differs from that of the
standard DP and CSL in Sec.~\ref{heating}. We use both forms in Eq.~\eqref{Lok1} of 
the Lindblad generator in the dissipator in Eq.~\eqref{compact.master}, to yield 
\begin{widetext}
\beq\label{clkin}
\frac{\D\rho(\pvo)}{\D t}=\frac{1}{\hbar^2}\int \frac{\D\kv}{(2\pi)^3} D_\kv
\left(
\left(m-\frac{\be\hbar^2 k^2}{8}+\frac{\be\hbar\kv}{4}\pvo\right)^2\rho(\pvo-\hbar\kv)
-\left(m+\frac{\be\hbar^2 k^2}{8}+\frac{\be\hbar\kv}{4}\pvo\right)^2\rho(\pvo)
\right).
\eeq
\end{widetext}
This is equivalent with a classical kinetic equation, therefore $\pv$ instead of $\pvo$ can be
written.
According to this kinetic equation, the momentum  jumps like 
$\pv\rightarrow\pv+\hbar\kv$ at probability rate 
\begin{equation}\label{clmomtransf}
\frac{m^2}{\hbar^2}\frac{\D\kv}{(2\pi)^3}D_\kv\left(1+\frac{\be}{8m}[\pv^2-(\pv-\hbar\kv)^2]\right)^2.
\end{equation}
This jump rate, unlike in standard DP and CSL, is not isotropic, and the anisotropy can generate 
the desired friction. Nevertheless, the above rate is subtle. Consider for simplicity a
momentum transfer of $\hbar\kv=\mp\kappa\pv$ where $0\leq\kappa\leq1$, where upper and lower  signs correspond to damping and heating  respectively. 
The difference between damping and heating rates is proportional to the following
expression:
\bea
& \left(1-\frac{\be \hbar^2 k^2}{8m}+\kappa\frac{\be p^2}{4m}\right)^2
-\left(1-\frac{\be \hbar^2 k^2}{8m}-\kappa\frac{\be p^2}{4m}\right)^2\nonumber\\
&=\kappa\frac{\be p^2}{m}\left(1-\frac{\be \hbar^2 k^2}{8m}\right).  
\eea
Damping dominates as long as $\be (\hbar^2 k^2/8m)<1$ and heating
takes over otherwise. Earlier, when we  defined $\Esi$ in Eq.~\eqref{Esi},
we noticed that the range of $k$ is $1/\si$, hence  $(\hbar^2 k^2/4m)\sim\Esi$.
Accordingly for damping, the largest value of $1/\be$ is about $2\Esi$. 
The forthcoming
analytic calculation shows that, indeed, there is an exact critical value of $1/ \be$
above which dissipation gives way to heating. 

In order to prove that the model exhibits the expected dissipative mechanism described in Eq.~\eqref{balance} for $\braket\Ho=\braket{\pvo^2/2m}$, 
we  derive the time-derivative of $\braket{\pvo^2}$. Since the dynamics of
$\pvo$ is semi-classical, we can derive the time-derivative of the equivalent semi-classical
mean $\braket{\pv^2}$.  According to the above discussion, the expression 
in Eq.~\eqref{clmomtransf} is the rate of random jumps $\pv\to\pv+\hbar\kv$, each of which leads to a change of $\hbar^2\kv^2+2\hbar\kv\pv$ for $\pv^2$. Hence, we can write 
\bea\label{dotpv2}
\frac{\D}{\D t}{\braket{\pv^2}}
&=&\frac{m^2}{\hbar^2}\int\frac{\D\kv}{(2\pi)^3}\int\D\pv D_\kv\rho(\pv)\times\\
&&(\hbar^2\kv^2+2\hbar\pv\kv)
\left(1+\frac{\be}{8m}[\pv^2-(\pv-\hbar\kv)^2]\right)^2\nonumber.
\eea
We temporarily  set $\hbar=1$ and introduce $a^2=(\be/4m)$.  We then rewrite the following expression 
\bea
&&(\kv^2+2\pv\kv)\left(1-\tfrac12 a^2\kv^2-a^2\kv\pv\right)^2\\
&=&k^2(1-a^2k^2+\tfrac14 a^4k^4)+a^2(3a^2 k^2-4)(\kv\pv)^2+...\nonumber\\
\label{eq.substitution}&\Rightarrow&k^2(1-a^2k^2+\tfrac14 a^4k^4)+a^2 k^2 (a^2 k^2-\tfrac43)p^2,
\eea
where the ellipsis stands for odd powers of $\kv$ to be cancelled when intergrating,
and we replaced $(\kv\pv)^2$ by $\tfrac13 k^2p^2$ also because of the isotropy of $D_\kv$
in the integral.  We insert in Eq.~\eqref{dotpv2}  the bottom line of the above 
expansion and perform the integral in $\pv$. We get the balance equation~\eqref{balance}, where
the power $P$ and the dissipation $\Gamma$ rates are respectively
\bea\label{PandGamma}
P
&=&\frac{m}{2}\int \frac{\D\kv}{(2\pi)^3}D_\kv 
k^2\left(1-\frac{\be\hbar^2}{4m}k^2+\frac{\be^2\hbar^4}{64 m^2}k^4\right),
\nonumber\\
\Gamma
&=&\frac{\beta m}{3}\int \frac{\D\kv}{(2\pi)^3}D_\kv 
k^2\left(1-\frac{3\be\hbar^2}{16m}k^2\right),
\eea
where we have restored $\hbar$ and $a^2=\be/4m$. The obtained result  shows that the energy of the system is dissipated, as expected, through the anisotropic process described by Eq.~\eqref{clmomtransf}. Moreover, we show in Appendix~\ref{app.1}, that such a result is independent of the specific form of the state $\hat \rho$ and is valid beyond the assumption of having $\hat \rho=\rho(\pvo)$.

The integrals in Eq.~\eqref{PandGamma} can be calculated analytically for the two models. They respectively read
\begin{equation}
\label{alphagammaDP}
    \begin{aligned}
P^\text{\tiny DP}  &= \frac{\hbar mG}{4\sqrt{\pi}\sigma^3}\left(1-\frac{3}{4}x_{\beta}^{2}+\frac{15}{64}x_{\beta}^{4}\right),\\
\Gamma^\text{\tiny DP}  &=\beta\frac{\hbar mG}{6\sqrt{\pi}\sigma^3}
\left(1-\frac{9}{16}x_\beta^2\right).
    \end{aligned}
\end{equation}
and
\begin{equation}
\label{alphagammaCSL}
    \begin{aligned}
P^\text{\tiny CSL}  &=\frac{3 m \gamma \hbar^2}{32 \pi^{3/2}\sigma^5}\left( 1-\frac{5}{4}x^2_\beta+\frac{35}{64}x^4_\beta \right), \\
\Gamma^\text{\tiny CSL}  &=\beta\frac{ m \gamma \hbar^2}{16 \pi^{3/2}\sigma^5}\left( 1-\frac{15}{16}x^2_\beta \right),
    \end{aligned}
\end{equation}
where we defined the dimensionless parameter
\beq
x_{\beta}^{2} = 2\be \Esi = \frac{\hbar^{2}\beta}{2m\sigma^{2}},
\eeq
which is the ratio of the elementary energy transfer $\Esi$ defined in Eq.~\eqref{Esi} to $1/(2\beta)$,
the latter  being  the equilibrium thermal kinetic energy at high temperatures.
Note that, according to the mapping in Eq.~\eqref{DrhoDPvCSL}, the relation 
 $P^\text{\tiny CSL}=-(\hbar\gamma/4\pi G)\partial P^\text{\tiny DP}/\partial(\si^2)$
holds, and similarly between 
$\Gamma^\text{\tiny CSL}$ and $\Gamma^\text{\tiny DP}$.  
The dissipative rates $\Gamma$ become negative
if $x_\be^2$, which is proportional to  the parameter $\be$, is larger than a critical value, which is different for the two models.

It is in order now to interpret our results, shown in Eqs.~\eqref{PandGamma}--\eqref{alphagammaCSL},
that exhibit the dissipation mechanism postulated by the balance Eq.~\eqref{balance}.
The equilibrium energy is obtained as $\braket{\hat{H}}_{\infty} =P/\Gamma$.
Following the equipartition theorem,
we define the effective temperature as $T=\tfrac23\braket{\hat{H}}_{\infty}/\kB$ 
and we introduce the parameter $\Tb=1/\be$ in place of $\be$. Then, we have
$x_\be^2=2\Esi/\Tb$, and we  can express the effective temperatures
of the two models as
\beq
\label{TeffDP}
T^\text{\tiny DP} 
=\Tb\frac{1-\frac{3}{2}(\Esi/\kB\Tb)+\frac{15}{16}(\Esi/\kB\Tb)^2}
{1-\frac{9}{8}(\Esi/\kB\Tb)},
\eeq
and 
\beq
\label{TeffCSL}
T^\text{\tiny CSL} 
=\Tb\frac{1-\frac{5}{2}(\Esi/\kB\Tb)+\frac{35}{16}(\Esi/\kB\Tb)^2}
{1-\frac{15}{8}\Esi/\Tb}.
\eeq
In the  regime $\kB\Tb\gg \Esi$,  the effective temperature $T$  asymptotically coincides with the parameter $\Tb$, which justifies our choice of parametrizing the dissipator in Eq.~\eqref{Ddiss} by $\be=1/\Tb$. 
When lowering the parameter temperature $\Tb$,
the effective temperature $T$ is also lowering.  But the dissipation rate $\Gamma$ 
is also decreasing and at a point the effective $T$ is no longer lowering together
with $\Tb$, but it is growing again and becomes infinite when the dissipation 
rate reduces to zero, i.e.~at $\kB\Tb=(9/8)\Esi$ in the DP model and at $\kB\Tb=(15/8)\Esi$ in the CSL model.
Below these critical temperatures, the negative dissipative rate $\Gamma$  is 
contributing to a higher heating power $P$ rather than balancing it. 
This effect follows from what we noticed about  the subtlety of momentum 
jump rate in Eq.~\eqref{clmomtransf}.

The standard DP and CSL models correspond to $T=\Tb=\infty$, where dissipative rates
$\Gamma$ vanish, the powers $P$ reduce to the expressions in Eq.~\eqref{PDP} and Eq.~\eqref{PCSL}, respectively. The kinetic energy ${\braket{\Ho}_t}$  goes to infinity
and, from a theoretical viewpont, this looks unphysical. In practice, however,
we face a different situation. The predicted powers  in Eq.~\eqref{PDP} and Eq.~\eqref{PCSL}
are extremely small and are typically masked by the environmental effects (see experimental investigations summarised in \cite{NatPhysMatteo}). 
Clearly, studying experimentally a system under the action of a collapse mechanism, but otherwise isolated, is impossible. Indeed, there will be always a coupling of the system with its surrounding environment. This might be the residual gas in the vacuum chamber, the blackbody radiation or the noises (e.g., seismic or electronic) that shake, and thus heat, the experiment (see for instance \cite{Cantilever-improved, cantilever-multilayer}). As a matter of fact, current laboratory efforts of isolation are
not yet able to exclude the values $T=\Tb=\infty$ neither for the DP or the CSL model,
however unphysical they would theoretically be. 

For this reason, we now consider the case of a system undergoing simultaneously to the dissipative collapse mechanism and the interaction of an external thermal environment. 
Let $T_\text{\tiny E}$ be  the temperature of the environment, and
let us define the power $P_\text{\tiny E}$ and the dissipative rate $\Gamma_\text{\tiny E}$
model the environmental effect on our particle, 
where  $\tfrac32\kB T_\text{\tiny E}=P_\text{\tiny E}/\Gamma_\text{\tiny E}$ is satisfied.
One can straightforwardly derive the evolution of the mean energy of the system, which reads:
\beq
\frac{\D}{\D t}\braket{H}_{t}=P + P_\text{\tiny E} -(\Gamma + \Gamma_{E})\braket{H}_{t},
\eeq
where $P$ and $\Gamma$ are those defined in Eq.~\eqref{PandGamma}. 
Consequently, the asymptotic mean energy is
\beq
\braket{\hat{H}}_{\infty} =\frac{P + P_\text{\tiny E}}{\Gamma + \Gamma_\text{\tiny E}}.
\eeq
According to the equipartition theorem, the asymptotic (equilibrium)
temperature of our particle is $T_\text{eff}=\tfrac23\kB^{-1}{\braket{\Ho}_\infty}$, i.e.:
\beq\label{eq.temp.eff}
T_\text{eff} =\frac{\Gamma T+\Gamma_\text{\tiny E}T_\text{\tiny E}}{\Gamma+\Gamma_\text{\tiny E}},
\eeq
where 
$T=\tfrac23\kB^{-1}P/\Gamma$ is the effective temperature obtained in Eqs.~\eqref{TeffDP}--\eqref{TeffCSL} of the collapse noise and 
$T_\text{\tiny E}=\tfrac23\kB^{-1}P_\text{\tiny E}/\Gamma_\text{\tiny E}$ is that of the thermal environment. The relation in Eq.~\eqref{eq.temp.eff} is fundamental when it comes to experiments. Indeed, it provides the experimental requirement  to be reached in terms of $\Gamma_\text{\tiny E}$ and $T_\text{\tiny E}$ to be able to measure the temperature $T$ of the collapse noise.

\section{Conclusion}
We introduced a simple and universal dissipative extension of the DP and the CSL models. Contrary to previous attempts \cite{smirne2015dissipative,BahSmiBas14}, our model modifies the collapse operator by adding (instead of a multiplying)
a new term leading to dissipative effects. A similar method has been considered for a different gravity-related model in \cite{DiBartolomeo}. Such a term is proportional to the divergence of the current and  is parametrized by the constant $\beta$ [cf.~Eq.~\eqref{Lr}]. 
We demonstrate that the model dissipates the current and  leads to the thermalisation of the system's energy to the asymptotic value of $\braket{\hat H}_\infty=\tfrac{3}{2}\kB T$, where the expression for $T$ is given in Eq.~\eqref{TeffDP} for the DP model and in Eq.~\eqref{TeffCSL} for the CSL model. 

We find a threshold temperature $T_0$, which is defined as
\beq
T_0=\frac{\hbar^2}{mk_\text{\tiny B}\si^2},
\eeq 
and is determined by the cutoff-length $\si$ of the DP and the CSL models. For $k_\text{\tiny B}/\beta$ much higher than $T_0$, the system's mean energy asymptotically converges to  $1/(2\beta)$, which suggests that $\beta$ can be interpreted as the inverse temperature of the collapse noise. Nevertheless, for generic values of $\beta$, the latter enters non-trivially in $\braket{\hat H}_\infty$.
The noise temperature $T$ becomes different from $\kB/\beta$ when the latter approaches $T_0$ from above. At a certain point, the
noise temperature $T$ inverts its trend with respect to $\kB/\beta$ and increases to infinity at $T_0$.
It is thus impossible to draw a clear one to one connection between the temperature $T$ of the collapse noise and the parameter $\beta$.
In general, the temperature of the collapse noise does not coincide with $\beta^{-1}/\kB$, and the latter plays the role of a parameter in the master equation that is detached from its familiar statistical mechanical interpretation.

\section*{Acknowledgments}
GDB acknowledges the financial support from University of Trieste and INFN.
MC is supported by UK
EPSRC (Grant No.~EP/T028106/1), the EU EIC Pathfinder project QuCoM (10032223) and PNRR PE National Quantum Science and Technology Institute (PE0000023). 
KP acknowledges support from the Centro Ricerche Enrico Fermi - Museo Storico della Fisica e Centro Studi e Ricerche “Enrico Fermi” (Open Problems in Quantum Mechanics project).
CC thanks to the INFN for supporting the research presented in this article.
LD was supported by the National Research, Development and Innovation
Office, Hungary, “Frontline” Research Excellence Programme grant
No.~KKP133827 and research grant No.~K12435.
KP, CC and LD acknowledge the support of Grant 62099 (QUBO Project) from the John Templeton Foundation. The opinions expressed in this publication are those of the authors and do not necessarily reflect the views of the John Templeton Foundation. CC, MD and LD  acknowledge support from the Foundational Questions Institute and Fetzer Franklin Fund, a donor advised fund of Silicon Valley Community Foundation, in particular the ICON project, (Grants No.~FQXi-RFP-CPW-2008 and  FQXi-MGA-2102).

\onecolumngrid

\appendix

\section{Single particle dissipative mechanism in the Heisenberg picture}
\label{app.1}

We re-derive here the results appearing in Sections \ref{DPCSLdiss} and \ref{single} in Heisenberg picture for a generic state $\hat \rho$. In particular, we show that the second term of $\mathcal D$ in Eq.~\eqref{Ddiss} leads to the dissipation of the current, and that the energy follows the balance equation displayed in Eq.~\eqref{balance}.
For the sake of simplicity, 
we focus on the case of a single particle. In such a case, the explicit forms of the Fourier transform of the mass density $\rodk$ and the current $\Jvok$, are
\bea\label{eq.def.rhok}
\rodk&=&m e^{i\kv\xvo},\qquad\Jvok=\frac{1}{2}\{\pvo, e^{i\kv\xvo}\}.
\eea
From these and from Eq.~\eqref{Weyl}, one can compute their commutator and anticommutator, which respectively read
\beq\label{eq.comm.anti}
[ \rod_{\q},\Jvok]=-m\hbar \q e^{i(\kv+\q)\xvo}=-\hbar \q \rod_{\kv+\q}\qquad \text{and}\qquad\{ \rod_{\q},\Jvok\}=me^{i(\kv+\q)\xvo}(2\pvo+\hbar(\kv+\q))=2m\Jvo_{\kv+\q}.
\eeq

The dynamics of the current due to the second term in Eq.~\eqref{Ddiss} is given, equivalentely, by the following two expressions
\begin{subequations}
\bea
{\dot\Jvo(\rv)}|_2&=&- \frac{i\be}{4\hbar}\int\!\!\int\!\! \D\sv \D\sv'\, D(\sv-\sv')\{\nabla_\sv\Jvo(\sv),[\rod(\sv'),\Jvo(\rv)]\},\\
&=&-\frac{\beta}{4\hbar}\int\frac{\D \kv'}{(2\pi)^3}e^{-i\kv'\rv}\int \frac{\D\kv}{(2\pi)^{3}} D_{\kv}\{\kv\Jvo_{\kv},[\rod_{-\kv},\Jvo_{\kv'}]\}.
\eea
\end{subequations}
By employing the second of these expressions, and merging it with Eq.~\eqref{eq.comm.anti}, one straightforwardly finds
\beq
{\dot\Jvo(\rv)}|_2=-\eta{\Jvo(\rv)},
\eeq
where
\beq
\eta=\frac{\beta m}{2}\int\frac{\D\kv}{(2\pi)^3}D_\kv \kv^2,
\eeq
{and shows explicitly that the contributions due $\hat {\bf Y}(\rv)$ in Eq.~\eqref{Jdot} vanishes exactly in the single fermion case.}
Similarly, one computes the following commutators of $\rodk$ and $\Jvok$ with $\hat H=\frac{\pvo^2}{2m}$:
\beq
[\rodk,\hat H]=-\frac{e^{i\kv\xvo}}{2}(\hbar^2k^2+2\hbar\kv\pvo)\qquad\text{and}\qquad
\left[\kv\Jvok,\hat H\right]=-\frac{\hbar e^{i\kv\xvo}}{4m}(\hbar k^2+2\kv\pvo)^2,
\eeq
from which one obtains
\bea
[\rod_{-\kv},[\rodk,\hat H]]&=&-m\hbar^2k^2,\\
\{\kv\Jvok,[\rod_{-\kv},\hat H]\}&=&2\hbar (\kv\pvo)^2+\frac{\hbar^3k^4}{2},\\
\left[\kv \Jvo_{-\kv},\left[\kv\Jvok,\hat H\right]\right]&=&-\frac{3\hbar^2k^2(\kv\pvo)^2}{m}-\frac{\hbar^4k^6}{4m}.
\eea
Owning that, for any spherically symmetric kernel $D_\kv=D_k$, the following holds
\beq
\int\D\kv D_k(\kv\pvo)^2=\int\D\kv D_k(k \hat p \cos\theta)^2
=\hat p^2\frac{4\pi}{3}\int\D k D_k k^4 =\int\D\kv D_k \left(\frac{k^2\hat p^2}{3}\right),
\eeq
we can substitute $(\kv\pvo)^2$ with $(k^2\hat p^2)/3$ [this has been used also in Eq.~\eqref{eq.substitution}]. Then, from Eq.~\eqref{Heis} we obtain the dynamics for the Hamiltonian:
\beq
\Dc^\dag \hat H=P-\Gamma \hat H,
\eeq
where the explicit form of $P$ and $\Gamma$ is given in Eq.~\eqref{PandGamma}. We underline that such an equation is state independent, thus it can be straightforwardly used to evaluate the expectation value of the energy for any state, also beyond the assumption of $\hat \rho= \rho(\pvo)$, which has been considered in the main text.

\end{document}